\documentclass[sigplan]{acmart}
\usepackage{booktabs} 

\usepackage{subfig}
\usepackage{float}
\usepackage{multirow, graphicx}
\usepackage{xspace}
\usepackage{hyperref} 
\usepackage{url}
\usepackage{listings}
\setcopyright{rightsretained}

\copyrightyear{2017} 
\acmYear{2017} 
\setcopyright{acmcopyright}
\acmConference[SysTEX'17]{SysTEX'17:2nd Workshop on System Software for Trusted Execution }{October 28, 2017}{Shanghai, China}
\acmPrice{15.00}
\acmDOI{10.1145/3152701.3152708}
\acmISBN{978-1-4503-5097-6/17/10}

\begin{document}
\title{Strongly Secure and Efficient Data Shuffle on Hardware Enclaves}

\author{Ju Chen}
\orcid{1234-5678-9012}
\affiliation{%
  \institution{Syracuse University}
  \streetaddress{}
  \city{Syracuse} 
  \state{New York} 
  \postcode{13244-1200}
}
\email{jchen133@syr.edu}

\author{Yuzhe (Richard) Tang}
\affiliation{%
  \institution{Syracuse University}
  \streetaddress{}
  \city{Syracuse} 
  \state{New York} 
  \postcode{13244-1200}
}
\email{ytang100@syr.edu}

\author{Hao Zhou}
\affiliation{%
  \institution{Syracuse University}
  \streetaddress{}
  \city{Syracuse} 
  \state{New York} 
  \postcode{13244-1200}
}
\email{hzhou09@syr.edu}

\renewcommand{\shortauthors}{Ju Chen, Yuzhe (Richard) Tang, Hao Zhou}

\begin{abstract}
Mitigating memory-access attacks on the Intel SGX architecture is an important and open research problem.
A natural notion of the mitigation is cache-miss obliviousness which requires the cache-misses emitted during an enclave execution are oblivious to sensitive data. 
This work realizes the cache-miss obliviousness for the computation of data shuffling.
The proposed approach is to software-engineer the oblivious algorithm of Melbourne shuffle~\cite{DBLP:conf/icalp/OhrimenkoGTU14} on the Intel SGX/TSX architecture, where 
the Transaction Synchronization eXtension (TSX) is (ab)used to detect the occurrence of cache misses.
In the system building, we propose software techniques to prefetch memory data prior to the TSX transaction to defend the physical bus-tapping attacks.
Our evaluation based on real implementation shows that our system achieves superior performance and lower transaction abort rate than the related work in the existing literature.
\end{abstract}



\maketitle

\providecommand{\memsetup}{\textsc{memsetup}\xspace}
\providecommand{\txbegin}{\textsc{txbegin}\xspace}
\providecommand{\applogic}{\textsc{applogic}\xspace}
\providecommand{\txend}{\textsc{txend}\xspace}
\providecommand{\checkcachedata}{\textsc{checkcachedata}\xspace}
\providecommand{\cmos}{\textsc{CMOS}\xspace}

\section{Introduction}

Today we witness the emergence of hardware enclaves, a trusted execution environment that protects trusted user program against the untrusted operating system. A notable example is the recently released Intel Software Guard eXtension (SGX~\cite{me:intelsgx})
with increasing adoption for secure public-cloud computing (e.g. in Microsoft Azure~\cite{me:mssgx} and Google Cloud Platform~\cite{me:googlesgx}).
Various side-channel attacks on hardware enclave exploiting memory access pattern~\cite{DBLP:conf/sp/XuCP15,DBLP:conf/uss/BulckWKPS17,DBLP:conf/uss/0001SGKKP17} have been proposed and demonstrated feasible in practice. Defending side-channel attacks on SGX-alike enclave architecture becomes an important and open research problem.

A natural notion of the defense is cache-miss obliviousness: The hardware enclave features a trusted processor issuing cache misses to access the memory in the untrusted world. Security can be assured by making the boundary crossing of cache miss oblivious to the sensitive data. This is especially effective to defending the physical attack, e.g. by bus tapping~\cite{DBLP:journals/iacr/CostanD16} and software attacks, e.g. page-fault controlled side-channel attack~\cite{DBLP:conf/sp/XuCP15}.
In addition to the strong security, cache-miss obliviousness helps induce better performance as cache-miss oblivious algorithms have lower time complexity than the classic word-oblivious algorithms~\cite{DBLP:conf/nsdi/ZhengDBPGS17,DBLP:conf/sp/LiuWNHS15} (see \S~\ref{sec:eval:perf} for performance discussion).

This work realizes the cache-miss obliviousness for data shuffling computation. The data shuffling is a basic operation used in many analytical computations. Specifically, we consider the Melbourne shuffle algorithm~\cite{DBLP:conf/icalp/OhrimenkoGTU14} which divides the data accesses in a shuffle to 1) the oblivious ones to a large external storage and 2) the non-oblivious ones to a small internal storage. We map the internal storage to the cache inside trusted process and the external storage to untrusted world (e.g. memory and disk). By this means, the cache misses that only touch the untrusted memory are made oblivious and thus safe to be disclosed. 

When engineering the above mapping paradigm on SGX, one challenge is how to conceal the internal storage in cache with assured isolation. We leverage Intel Transaction Synchronization eXtension~\cite{me:inteltsx}, a Hardware Transaction Memory feature in the latest Intel Skylake processor. TSX was originally designed for efficient concurrency. In this work, 
we propose a technique to enable the TSX transaction to detect cache misses; the observation is that TSX provides the capability of detecting early cache write-back (before the transaction commit) and cache miss can be detected if one can {\it equate cache miss with cache write-back.} The proposed technique pre-fetches all memory referenced in a transaction and make them dirty cache lines so that their eviction triggering write-back can be detected. 

In addition to cache-miss detection, we propose techniques to {\it avoid} unnecessary transaction abort by carefully aligning data in memory and ensuring no conflict during data prefetch. Our technique leverages the specific semantic of oblivious computation and is distinct from the compiler-based partitioning schemes such as T-SGX~\cite{shih2017t}.

We conduct algorithmic analysis to demonstrate the better complexity of our hybrid external-oblivious algorithms. More importantly, we conduct performance study based on real implementation, with the hope of verifying the advantage of systems-level performance of the cache-miss oblivious computation. 
The systems-level performance study is necessary even in the presence of algorithmic analysis. Because the overhead induced by cache-miss obliviousness, mostly due to TSX transaction execution, is higher than that of word obliviousness. Our performance study in realistic settings show that the systems-level overhead of cache-miss obliviousness is negligible in the presence of better time complexity, and cache-miss obliviousness causes an overall better performance than the word-oblivious computation.

\section{Preliminary}
\subsection{Intel Software Guard eXtension (SGX)}
\label{sec:sgx}

Intel SGX is a security-oriented x86-64 ISA extension on the Intel Skylake CPU, released in 2016.
SGX provides a ``security-isolated world'' for trustworthy program execution on an otherwise untrusted hardware platform.
At the hardware level, the SGX secure world includes a tamper-proof SGX CPU which automatically encrypts memory pages (in the so-called enclave region) upon cache-line write-back. Instructions executed outside the SGX secure world that attempt to read/write enclave pages only get to see the ciphertext and can not succeed. 
SGX's trusted software includes only unprivileged program and excludes any OS kernel code, by explicitly prohibiting system services (e.g. system calls) inside an enclave.

To use the technology, a client initializes an enclave by uploading the in-enclave program and uses SGX's seal and attestation mechanism~\cite{Anati_innovativetechnology} to verify the correct setup of the execution environment (e.g. by a digest of enclave memory content). During the program execution, the enclave is entered and exited proactively (by SGX instructions, e.g. \texttt{EENTER} and \texttt{EEXIT}) or passively (by interrupts or traps). These world-switch events trigger the context saving/loading in both hardware and software levels.
Comparing prior TEE solutions~\cite{me:txt, me:tpm, me:tzone, me:scpu}, SGX uniquely supports multi-core concurrent execution, dynamic paging, and interrupted execution.

\subsection{Intel Transactional Synchronization eXtension (TSX)}
\label{sec:tsx}

The purpose of TSX is to enable atomic execution of a code block or transaction from other processors' view point. This goal entails two requirements: 1) During the transaction execution, it is fully contained inside a processor and its memory-access requests can all be resolved inside the data cache. In other words, the transaction writes are buffered by dirty cache lines without being reflected in the memory. 2) By the end of transaction execution, the cached writes are successfully written back and the transaction can be committed only when there is no data conflict with other processor. A conflicting data access occurs when there is a location being accessed by two processors and at least one access is a transactional write. 

To realize the two requirements, the TSX hardware supports the capability of aborting the execution of a transaction under various causes. It aborts a transaction when data conflict is detected at the transaction commit time (Abort Cause AC1). In order to detect data conflict, the hardware needs to track both the readset and writeset of a transaction. 
The writeset needs to be kept inside the L1 data cache (L1D) and readset needs to be inside the L3 cache. Thus, it aborts the transaction when the dirty data-cache lines are evicted, triggering cache write-back, before the end of transaction (AC2). It also aborts upon the readset exceeding the L3 cache (AC3). In addition, it aborts upon various systems events such as page-fault, interrupts and other exceptions delivered to the processor (AC4). 

While TSX is originally designed for the performance and programmability in multiprocessing, its capabilities can be (ab)used for security purposes: It can be used to realize the cache-based or register-based computation~\cite{DBLP:conf/ndss/GuanLLJ14} and to protect the private key from leaving a processor~\cite{DBLP:conf/sp/GuanLLJW15} by leveraging the TSX capability of detecting early cache write-back. T-SGX~\cite{shih2017t} defends the page-fault side-channel attacks by leveraging the TSX capability that page-fault events are intercepted by the TSX abort handlers before the untrusted OS. This work uses TSX for detecting cache-misses and for defending side-channel attacks. The use of TSX in this work is elaborated below.

\subsection{Use of TSX for Detecting Cache-Miss}
\label{sec:usetsx}
\begin{figure}
\centering
  \includegraphics[width=0.265\textwidth]{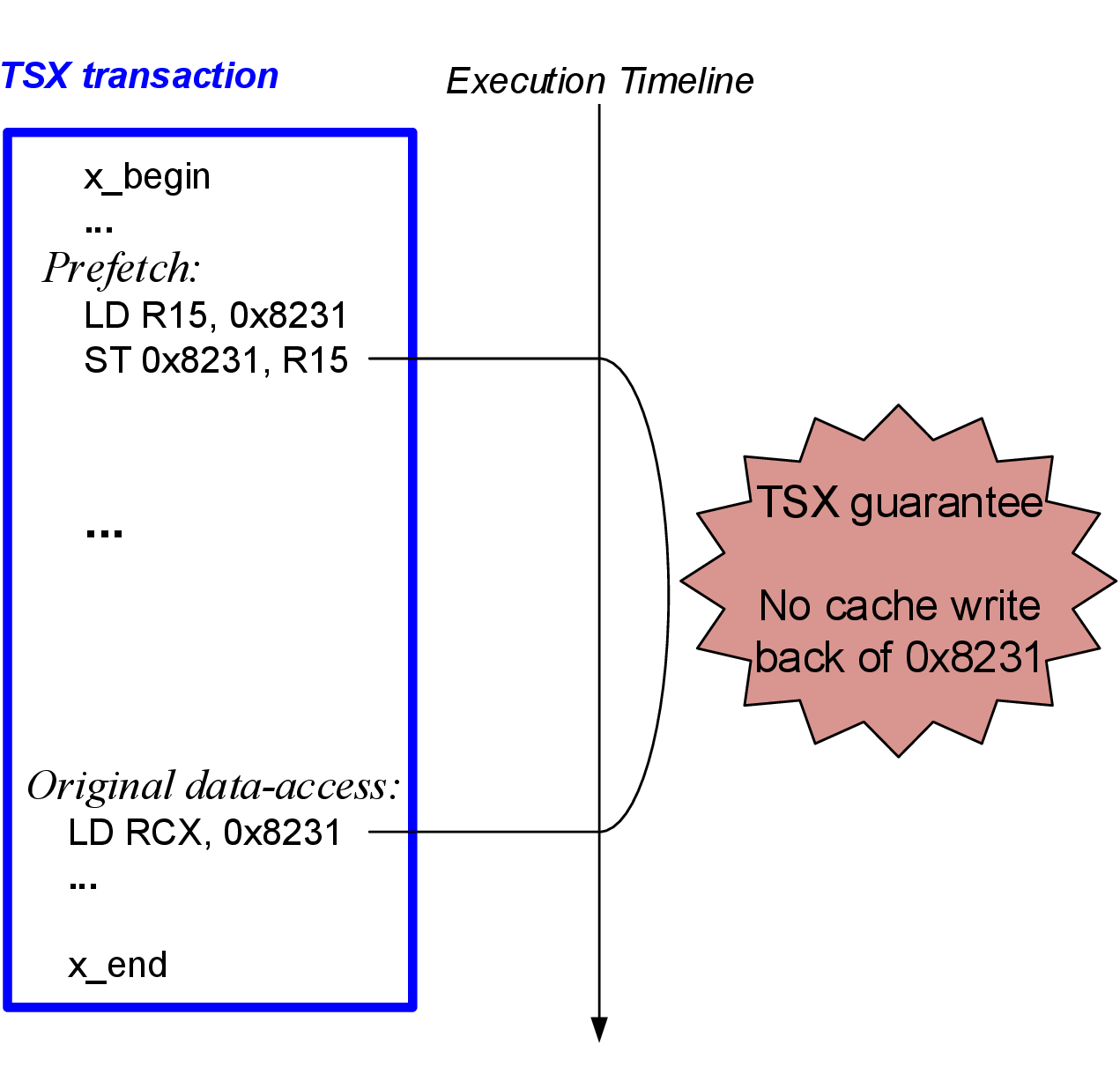}
\caption{Intel TSX (ab)used to detect cache misses: The hack here is that no cache write-back of a dirty line means the line stay present in the cache, implying a cache hit upon a memory request.
}
  \label{fig:ensurecachemiss}
\end{figure}

Our work requires to conceal the access-leaky internal storage in cache. Any access to the internal storage cannot be resolved by cache miss, which would otherwise leak the sensitive access. To conceal the internal storage in the last-level cache (LLC), it is equivalent to ensure no LLC cache miss during the time period when the internal storage is being accessed. 
Thus, the key requirement of our work is to ensure no LLC cache miss caused by internal-storage access.

TSX as is, however, does not provide this capability. For instance, a cache miss in a transaction does not necessarily cause the transaction to abort. A cache miss would abort a TSX transaction when it replaces a line that is read or written by the same transaction.
The key insight of this work is to prefetch the entire read/write set of a transaction, such that the actual access must be served by cache hits (without leaky cache miss). More concretely, TSX guarantees any attempt to replace the prefetched cache line would abort the transaction. With unreplace-able prefetched lines, the actual access is guaranteed to be served by cache hits..

For example, consider the memory reference request of \texttt{LD RCX,0x8231} in the code sample in Figure~\ref{fig:ensurecachemiss}. The memory reference can be resolved by a cache hit or a miss. To ensure no chance of cache miss, we prefetch the data to the LLC in the beginning of the transaction (i.e. ``\texttt{LD R15,0x8231}''). After this instruction, the LLC cache-line buffering the content at \texttt{0x8231} is recorded into the readset of this transaction and is ``pinned'' there; TSX guarantees any attempt to replace the LLC line will abort the transaction, which is further captured by TSX abort handler. In other words, if the transaction does not abort when the execution reaches instruction ``\texttt{LD RCX,0x8231}'', the prefetched cache line is still present at least in the LLC and the memory reference must not cause LLC miss or cause any traffic on the system bus.

In general, the capability of prefetching transaction read-/write-set and pinning them to unreplaceable cache-lines can assist mitigate various memory-access attacks including physical bus tapping and cache-timing attacks. Bus tapping can be mitigated due to pinned cache-line guarantees cache hits. The cache-timing attacks are mitigated due to sharing cache-lines between transactions.

\section{System Design and Impl.}

\subsection{Threat Model}
We mainly consider a memory-access attacker who either directly sniffs the out-of-process memory-access traffic (e.g. by bus tapping or by page-fault channel~\cite{DBLP:conf/sp/XuCP15}) or indirectly monitors the side-channel of cache timing~\cite{DBLP:conf/uss/BulckWKPS17}. 

The non-goals of this work includes the following attacks.
1) This work is complementary to rollback attacks load sealed but stale data across power cycles and restore the system to a stale state. 2) Given memory store both data and code, this work focuses on data-memory. The code-access attacks are orthogonal that exploit the access pattern to the memory region storing code, and that can be defended by existing techniques~\cite{DBLP:conf/asplos/LiuHMHTS15,shih2017t} on a small memory. 3) We don't consider other side-channel attacks exploiting timing information or power usage~\cite{DBLP:conf/uss/AlmeidaBBDE16}. 4) We don't consider denial-of-service attacks that the adversary declines to serve the requests from the enclave.

\subsection{Security Definition}

Intuitively, the memory-access obliviousness states that the memory access trace in a program execution is independent with any computation data (involving both input and intermediate data). Consider the execution of a program $P$ with data input $I$. The execution produces the memory-access trace $T$ that consists of all the last-level cache misses. The obliviousness requires that given two data values, $I_0$ and $I_1$, the an oblivious execution produces the same trace, that is, $T_P(I_0)=T_P(I_1)$. 
This definition assumes deterministic computation induced by $P$ and is about ``perfect'' obliviousness in the sense that it requires the traces under different input data stay exactly the same.
Due to the systems nature of this work, we skip the more formal and generic definition of cache-miss obliviousness (e.g. based on indistinguishability formation~\cite{DBLP:books/crc/KatzLindell2007}). 

\subsection{Step 1: Mapping Melbourne Shuffle to SGX}

\begin{figure*}
\centering
\subfloat[Melbourne shuffle: The example shows the shuffle of data array (a,b,c,...,i) based on permutation (3,1,6,5,7,2,0,8,4). The bucket used in the distribute pass is vertical and the bucket in the sort pass is horizontal]{
  \includegraphics[width=0.475\textwidth]{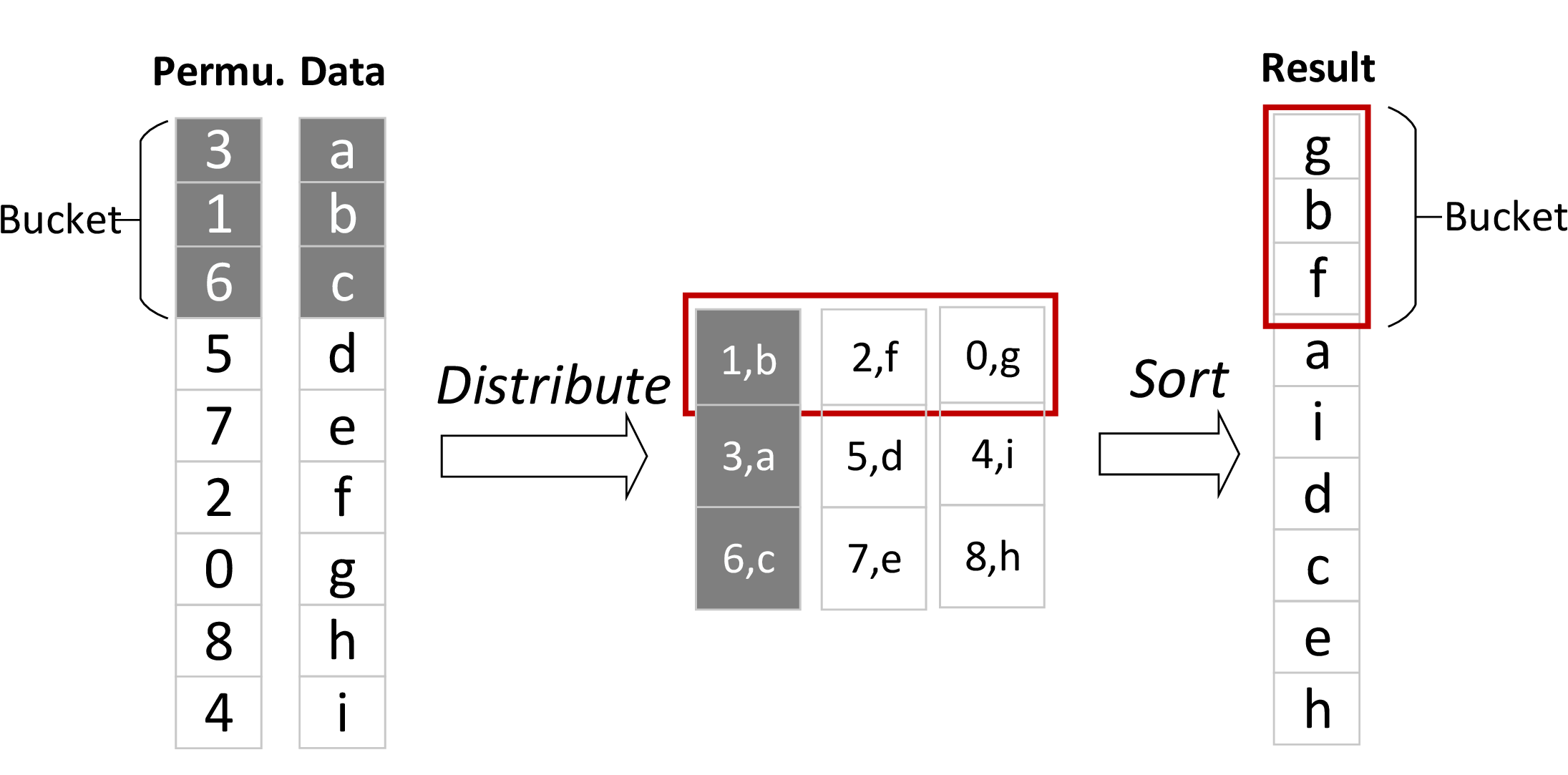}
  \label{fig:shuffle:melalgo}
}
\hspace{0.5in}
\subfloat[Melbourne shuffle run in SGX/TSX]{
  \includegraphics[width=0.25\textwidth]{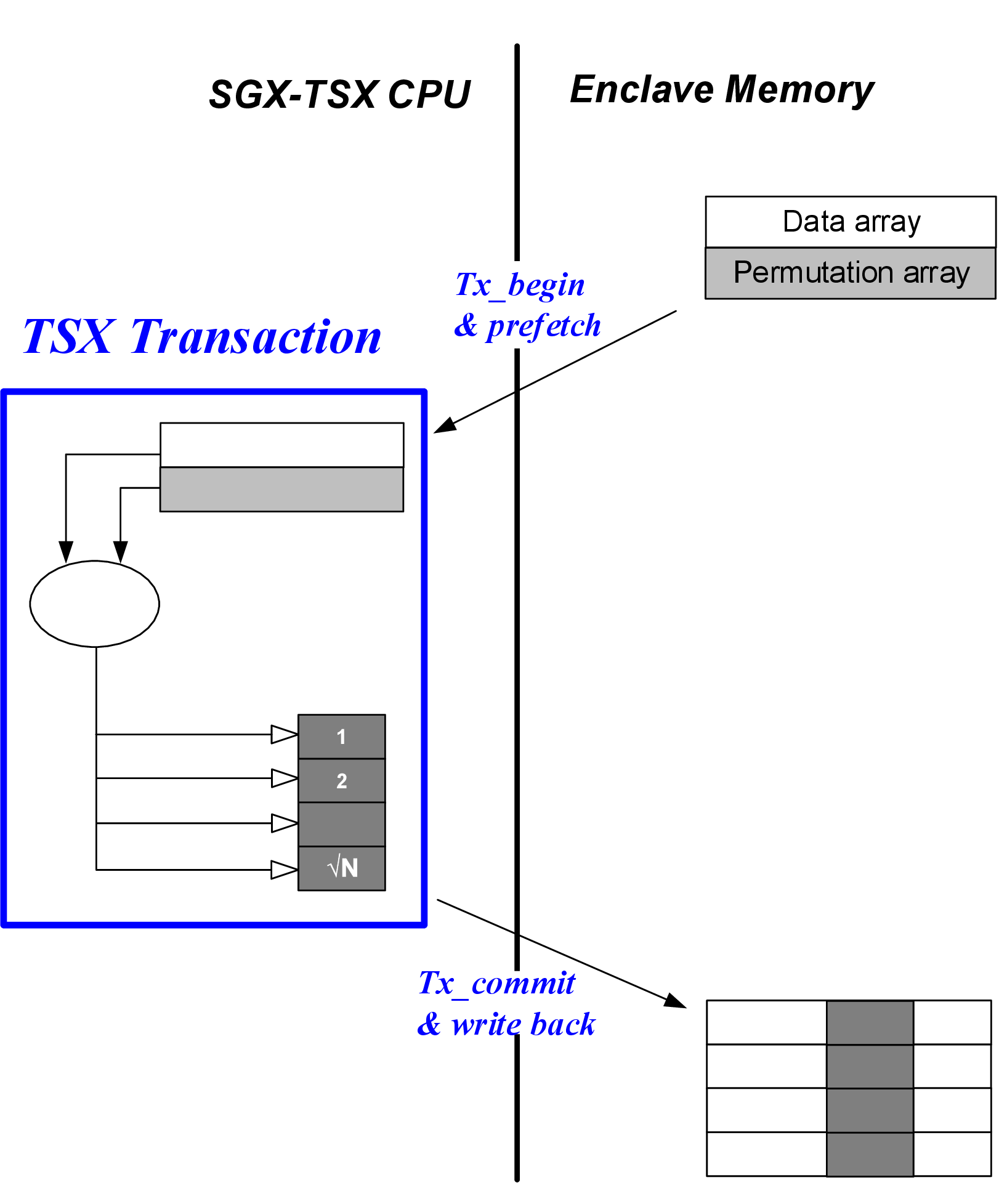}
  \label{fig:shuffle:mel}
}
\caption{Engineering Melbourne shuffle algorithms on SGX/TSX}
\end{figure*}

In this work, we focus on implementing cache-miss oblivious data shuffling. Data shuffle is a fundamental operator in oblivious data analysis computation. 
We describe the engineering of Melbourne shuffle~\cite{DBLP:conf/icalp/OhrimenkoGTU14} on Intel SGX. 
The idea is to map the program of Melbourne shuffle to TSX transactions and to isolate the leaky data access in cache by abusing TSX capability.
Note this work only considers single-threaded execution.

{\bf Preliminary of Melbourne shuffle~\cite{DBLP:conf/icalp/OhrimenkoGTU14}}:
Melbourne shuffle is an oblivious, randomized algorithm for data shuffle. Given a data array and a permutation (of the same length), the computation of a data shuffle produces an array that reorders the data array based on the permutation. Internally, the Melbourne shuffle works in two data scans or passes, where the first pass, called distribution, scans the data array at the granularity of size $\sqrt{N}$ buckets and reorders individual buckets non-obliviously with the size-$p\log{N}\sqrt{N}$ internal memory. The second pass scans the array and sorts reorganized buckets internally. Figure~\ref{fig:shuffle:melalgo} illustrates an example of Melbourne shuffle. Overall, the trusted memory is $\log{N}\sqrt{N}$ with array length $N$. The details can be found in the original paper~\cite{DBLP:conf/icalp/OhrimenkoGTU14}.

{
\begin{figure}[h!]
\lstset{ %
  backgroundcolor=\color{white},   
  basicstyle=\scriptsize\ttfamily,        
  breakatwhitespace=false,         
  breaklines=true,                 
  captionpos=b,                    
  commentstyle=\color{mygreen},    
  deletekeywords={...},            
  escapeinside={\%*}{*)},          
  extendedchars=true,              
  keepspaces=true,                 
  keywordstyle=\color{blue},       
  language=Java,                 
  morekeywords={*,...},            
  numbers=left,                    
  numbersep=5pt,                   
  numberstyle=\scriptsize\color{black}, 
  rulecolor=\color{black},         
  showspaces=false,                
  showstringspaces=false,          
  showtabs=false,                  
  stepnumber=1,                    
  stringstyle=\color{mymauve},     
  tabsize=2,                       
  title=\lstname,                  
  moredelim=[is][\bf]{*}{*},
}

\begin{lstlisting}
int[] Melbourne_shuffle(int[] data,
                        int[] perm){
  perm_r=gen_perm();
  data_r=shuffle_pass(data,perm_r);
  perm_rr=shuffle_pass(perm,perm_r);
  return shuffle_pass(data_r,perm_rr);
}
int[] shuffle_pass(int[] data,
                   int[] perm){
  int[][] inter=distribute(data,perm);
  return cleanup(inter);
}
int[][] distribute(int[] data,
                        int[] perm){
   for(int i=0;i<sqrt(length(data));i++){
      inter[i]=tx_bucket_perm(data,perm,i);
   }
   return inter;
}
int[] cleanup(int[][] inter){
   List res;
   for(i < sqrt(length(data))){
     res.add(tx_bucket_sort(inter,i));
   }
   return res.toarray();
}
\end{lstlisting}
\caption{Melbourne shuffle mapped to TSX transactions}
\label{lst:melshuffle:code}
\end{figure}
}

The Melbourne shuffle is mapped to TSX transactions such that the accesses to internal storage are kept inside transactions while external oblivious data accesses are kept outside transactions. In Melbourne shuffle, the mapping is illustrated in Figure~\ref{lst:melshuffle:code} where the bucket-wise permutation multiplication in the ``distribute'' pass and the bucket-wise sort in the ``cleanup'' pass are mapped to individual transactions. 

One implication of this mapping is that the internal storage of $\log{N}\sqrt{N}$ must be smaller than that of the size of L1 cache, which is one factor that constrains the scalability of \cmos on real SGX hardware (see \S~\ref{sec:eval:perf}).

\subsection{Step 2: Isolating Cached Data by TSX}

Isolating cached data is realized by data prefetching which simply prefetch all data referenced inside a transaction. Given a prefetched line, the TSX capability guarantees that at least the line will not be replaced from LLC during the transaction. Our goal in this work is to avoid self-eviction, that is, the dataset prefetched does not conflict each other. Here, we consider both conflict and capacity cache misses during prefetching. Given a set-associative cache, we lay out memory properly such that the number of conflicts in each cache set do not exceed the capacity (i.e. the number of ways). This applies for both L1 and LL caches.

Figure~\ref{fig:shuffle:mel} shows how isolation is realized with the distribution phase of Melbourne shuffle. First, data is prefetched from the enclave to the cache. Second, it runs non-oblivious computation on the cached data; this phase is wrapped in TSX transactions to ensure no cache miss. Third, the end of the transaction triggers the write-back of cached lines to the enclave memory.
The second-phase transaction, in particular, takes two continuous memory regions as input and output data stored in another contiguous memory region. To avoid conflict in this layout, we partition the three regions at granularity of cache lines and precompute (at compilation time) that the number of conflicts in each cache set does not exceeds the number of ways.

\subsection{Implementation Notes}

{\bf Abort Handling}:
Transaction aborts are handled by re-executing the transactions. Before entering the transaction, the context (all the values in registers) is saved to a memory area pointed by a reserved register \texttt{R15}. Upon aborts, the handler reloads the context prior to the transaction from \texttt{R15}, before jumping back to the beginning of the transaction to re-execute it. An important property is that the value of \texttt{R15} must be preserved through the regular transactional path and abort path. 

{\bf CPUID}: 
Obtaining cache information (e.g. cache sizes) is realized by calling the functions provided in Intel SGX SDK~\cite{me:intelsgx} which switches out to the untrusted world and calls \texttt{CPUID} instructions. Here, the result of \texttt{CPUID} is not necessarily be trusted. The enclave can run test program to evaluate the cache sizes itself. Concretely, it can read a series of arrays with increasing lengths and the maximal length without aborting transaction is the size of LLC cache. Similarly, it can writes to a series of length-increasing arrays to obtain the maximal length as L1 cache size.

{\bf Randomness generation}: Melbourne shuffle is a randomized algorithm and we generate true randomness using SDK-provided function \texttt{sgx\_read\_rand}.

\section{Evaluation}

This section evaluates the performance and abort rate of \cmos system. Specifically, it aims at answering the following two questions.

\begin{itemize}
\item
What is the performance of \cmos comparing the baseline of word-oblivious shuffle and pure transaction-based protection?
\item
What is the abort rate of \cmos comparing with the implementation without prefetching? 
\end{itemize}

\subsection{Performance}
\label{sec:eval:perf}

\begin{figure}
\centering
  \includegraphics[width=0.4\textwidth]{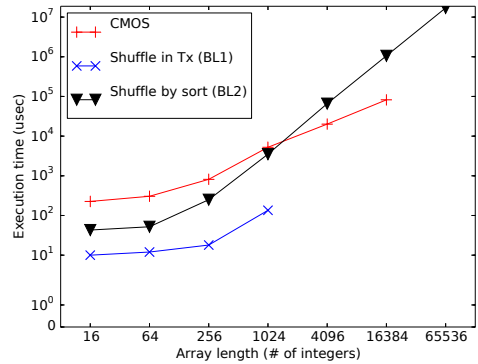}
\caption{Execution time of \cmos and other shuffle baselines}
\label{exp:shuffle:small}
\end{figure}

We consider two baselines for performance comparison against \cmos. 
The first baseline (BL1) is pack the naive, non-oblivious shuffle algorithm in TSX transactions whose time complexity at $O(N)$ is better than the $O(\sqrt{N}\log{N})$ complexity of Melbourne shuffle. 
The second baseline is the word-oblivious shuffle that is realized by running a bubble sort on the permutation array. This baseline has worse time complexity than \cmos, but it does not need run transactions, adding performance uncertainty.

{\bf Experiment setup}: We did all the experiments on a laptop with an Intel 8-core i7-6820HK CPU of 2.70GHz, 32KB L1 and 8MB LL cache, 32 GB RAM and 1 TB Disk. This is one of the Skylake CPUs equipped with both SGX and TSX features. 
We use numeric datasets and generate them randomly. 

In the evaluation, we measure the execution time and the maximal data size supported. In transaction-based approaches, the individual transaction is bounded by L1 cache size, which in our platform is $32$ KB. 

The performance result is illustrated in Figure~\ref{exp:shuffle:small}. It can be seen that BL1 is the most efficient but with limited scalability. BL2 has the best data scalability but may not be efficient, especially when data size is large. \cmos starts to show superior performance when data size roughly grows beyond $4096$. \cmos can scale to larger dataset than BL2 because of its smaller space complexity. It has smaller execution time than BL1 because of the better time complexity. 

\subsection{Transaction Abort Rate}
\label{sec:eval:util}

\begin{figure}
\centering
\subfloat[With Baseline 1,2]{
  \includegraphics[width=0.25\textwidth]{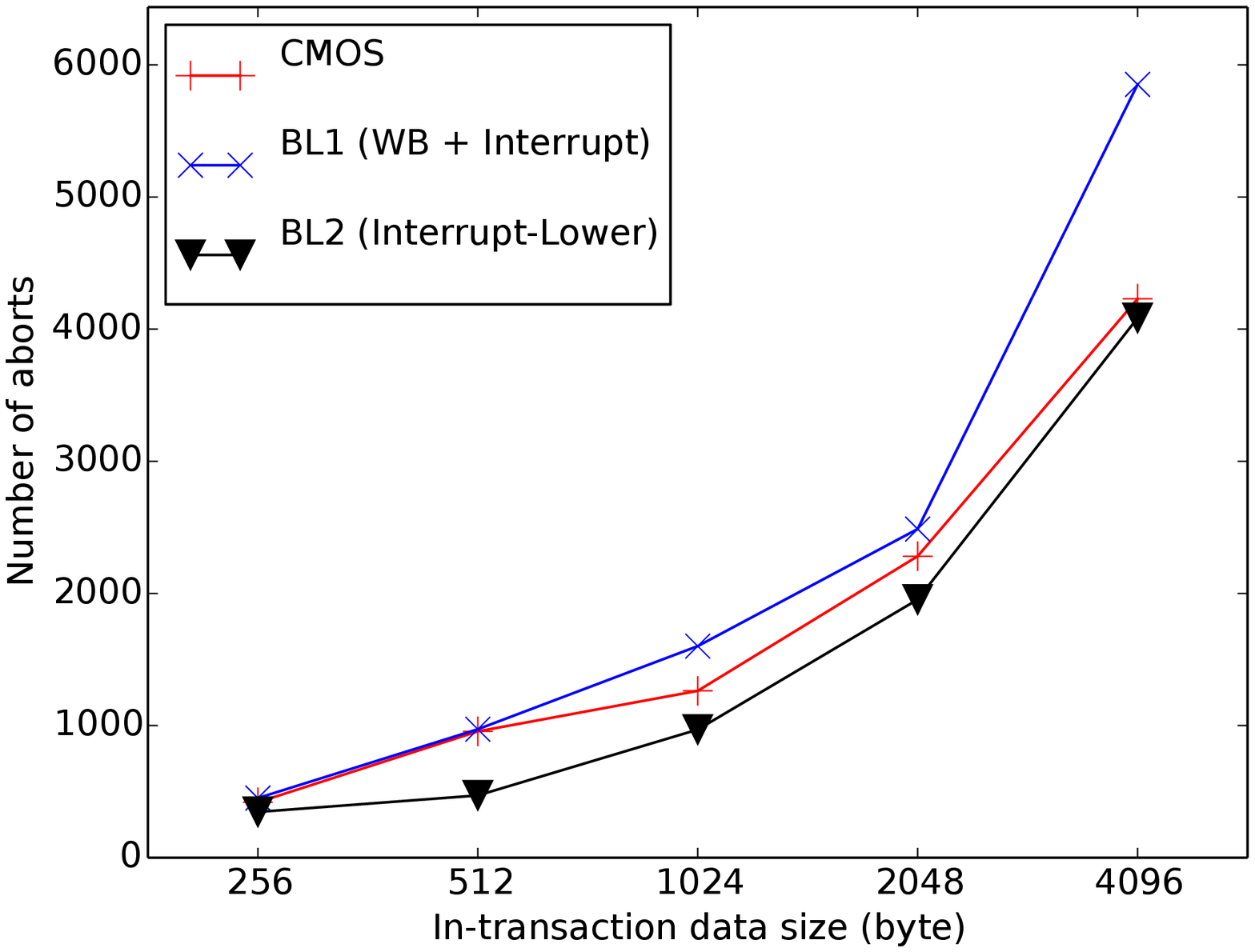}
\label{exp:aborts1}
}
\subfloat[With Baseline 3]{
  \includegraphics[width=0.25\textwidth]{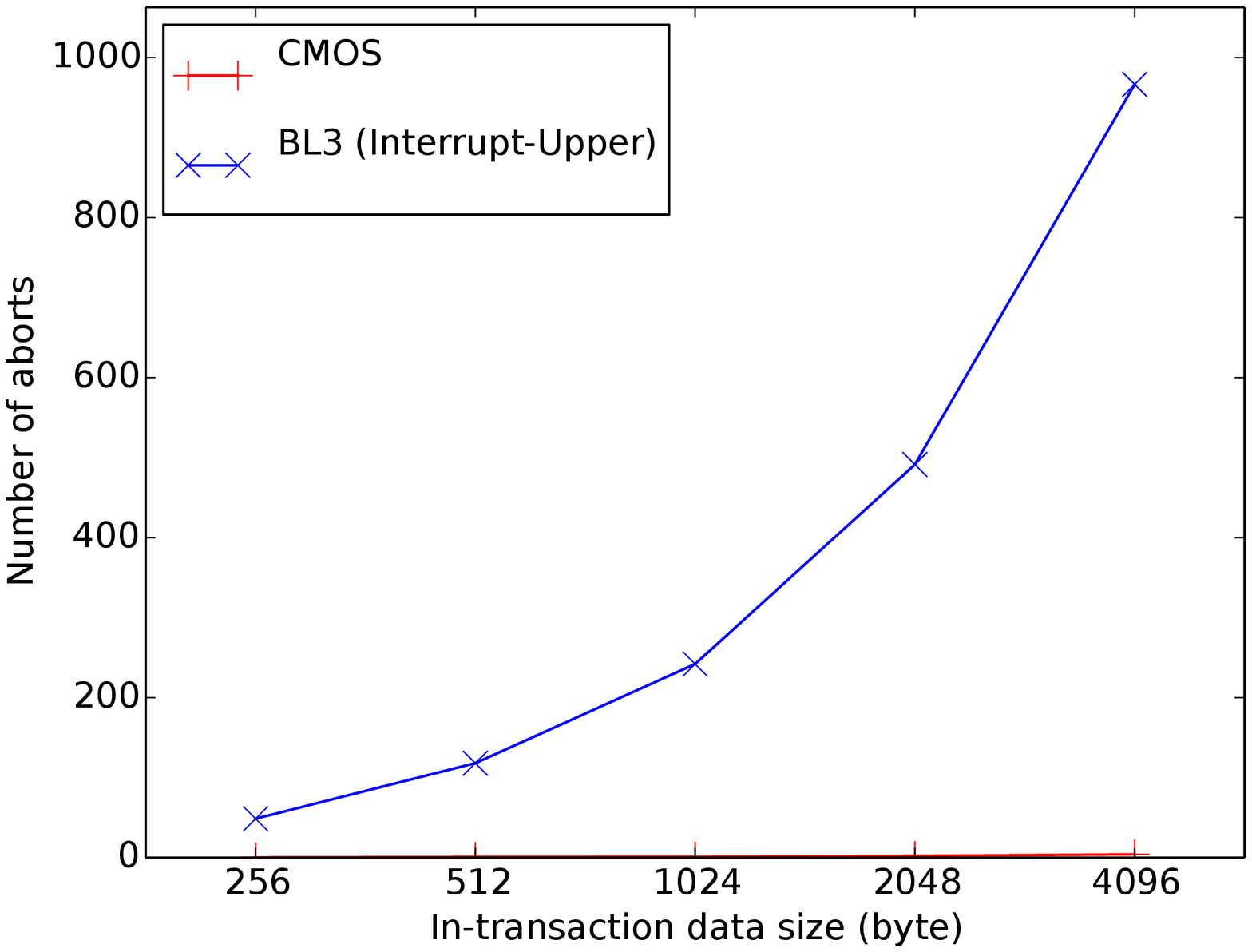}
\label{exp:aborts2}
}
\caption{Number of transaction aborts}
\label{exp:aborts}
\end{figure}

One of the design goals of prefetch in \cmos is to reduce the rate of self-abort. To evaluate the design effectiveness, we run the program several runs and measure the average abort rate. We consider the first baseline that simply skips the external prefetching. The baseline would cause transactions to abort due to both interrupt (AC4) and cache write-back (AC2) (recall \S~\ref{sec:tsx}). 
We also consider two additional baselines that do not cause write-back by placing \texttt{nop} instructions in a transaction. This way, the baselines are only caused by interrupt. We implement two variants: the second baseline, named interrupt-lower, runs enough \texttt{nop} instructions such that the running time is equal to the case that all memory references are solved by cache hits. The third baseline, named interrupt-upper, runs enough \texttt{nop} instructions such that the running time is equal to the case that all memory references are solved by cache misses.
Note that to determine the appropriate number of \texttt{nop} instructions, we profile our SGX hardware using some benchmark programs (e.g. calling cache-flush instructions to measure the cache-miss time). 

We conduct experiments by running the four approaches above multiple times. We vary the size of data referenced inside the transaction and report the percentage of transaction aborts during the execution, in Figure~\ref{exp:aborts}. As can be seen, \cmos causes aborts that are fewer than BL1, close to BL2, and are much fewer than BL3. Given BL2 is ideal lower bound of aborts, the result shows that the \cmos design of double prefetch is highly effective in reducing cache write-back aborts. In addition, \cmos is approximately the same with the Ideal-lower bound, which shows the \cmos design {\it eliminates} the cache write-back aborts as the ideal can not be aborted by cache write-back.

\section{Related Work}

In this section, we survey the related work on defending memory-access side-channel attacks on hardware enclaves. 

Making memory access oblivious is a feasible defense strategy to side-channel attacks in general. Comparing the cumbersome Oblivious RAM protocols~\cite{DBLP:journals/jacm/GoldreichO96,DBLP:conf/ccs/StefanovDSFRYD13}, oblivious algorithms~\cite{DBLP:conf/sp/LiuWNHS15} are more lightweight and are promising towards practical attack defense. Various data-analytical systems are developed based on the computation-specific oblivious algorithms, such as  Opaque~\cite{DBLP:conf/nsdi/ZhengDBPGS17} for relational data analytics and oblivious machine learning~\cite{DBLP:conf/uss/OhrimenkoSFMNVC16}. These systems instantiate the notion of word-obliviousness; recall that its goal is to make memory references at word granularity oblivious. ObliVM~\cite{DBLP:conf/sp/LiuWNHS15} is a source-to-source program transformation that translates the annotated program into efficient oblivious algorithms. This line of research does not consider external-oblivious algorithms and does not attain the cache-miss obliviousness on SGX.

T-SGX~\cite{shih2017t} takes a general-purpose approach to defend the page-fault side-channel attack~\cite{DBLP:conf/sp/XuCP15}. T-SGX's approach is to assume the allocated memory is large enough to hold the data referenced by the application, such that during execution there is no page-fault. It  leverages the TSX capability of detecting page-faults in user-space programs. It partitions the program and wraps the partitions into individual TSX transactions. Similarly, work~\cite{DBLP:conf/ccs/ShindeCNS16} takes a compiler approach to defend the page-fault based side-channel attacks. It is based on the notion of page-fault obliviousness.

This work is different from T-SGX in the following senses: 1) The goals are different: 
T-SGX is a defense of page-fault side-channel attacks, and this work is to defend all software/hardware memory-access pattern attacks on SGX. 
T-SGX makes page-fault in enclave unobservable, while this work is to make enclave execution cache-miss oblivious. 
2) The approaches are different: T-SGX defends by detecting page faults inside TSX transactions, while this work defends by detecting cache misses inside TSX transactions. While both seem to rely on TSX, their {\it use} of TSX transactions is fundamentally different, which will be elaborated on in the next paragraph.
3) The applicability is different: T-SGX is a general-purpose, compiler-based solution, and this work is specific to data analytics and leverage the corresponding ``semantics'' for better efficiency. In addition, T-SGX mainly focus on the security of code-page execution, while this work focuses on data-obliviousness.

The use of TSX transactions in T-SGX and this work is different. In T-SGX, it assumes a memory large enough to store all data pages and ensure no page-fault during enclave execution. Given the limited ``size'' allowed by a TSX transaction, T-SGX packages enclave computation in as many TSX transactions as needed and ensure the security of page-access across transactions by placing all the code outside transaction on a single page (so-called Springboard page). It is important to note the Springboard code-page does not access memory on any other pages. 
In this work, the TSX transactions are used to isolate the enclave computation inside the processor and to detect any (unexpected) cache misses. Across transactions, the data security is ensured by running external oblivious algorithms.
This work is applicable to a more realistic setting when handling a large volume of data, that is, we allow page-fault during program execution and the working-set memory can be much smaller than original dataset.

\section{Conclusion}
This work defends enclave side-channel attacks by cache-miss obliviousness. The proposed approach is software engineering the target oblivious computation on top of the SGX and TSX platform. It proposes cache-miss oblivious algorithms with small trusted space. It has several software-engineering strategies that package the computation into TSX transactions, achieving cache-miss obliviousness. Through initial evaluation, the performance overhead of cache-miss obliviousness is much smaller than that of the Strawman approach.

\section*{Acknowledge} This research is partly supported by the Cyber Research Institute in Rome, NY, under Grant Number \#28254.

{
\scriptsize
\bibliographystyle{ACM-Reference-Format}
\bibliography{yuzhetang,sgx,odb,latex,crypto} 
}
\end{document}